# Adsorption of Externally Stretched Two-Dimensional Flexible and Semi-flexible Polymers near an Attractive Wall

Pui-Man Lam and Yi Zhen
Physics Department, Southern University
Baton Rouge, Louisiana 70813

Haijun Zhou and Jie Zhou
Institute of Theoretical Physics, The Chinese Academy of Sciences
Beijing 100190, China

We study analytically a model of a two dimensional, partially directed, flexible or semiflexible polymer, attached to an attractive wall which is perpendicular to the preferred direction. In addition, the polymer is stretched by an externally applied force. We find that the wall has a dramatic effect on the polymer. For wall attraction $\varepsilon_1$ smaller than the non-sequential nearest neighbor attraction $\varepsilon$, the fraction of monomers at the wall is zero and the model is the same as that of a polymer without a wall. However, for $\varepsilon_1$ greater than $\varepsilon$, the fraction of monomers at the wall undergoes a first order transition from unity at low temperature and small force, to zero at higher temperatures and forces. We present phase diagram for this transition. Our results are confirmed by Monte-Carlo simulations.

## I. Introduction

The adsorption of polymers at a attractive wall is a well-studied problem from theoretical [1-7], computer simulation [7-10] and experimental [11,12] points of view. Due to the advancement of single-molecule force manipulation methods [13-15], the conformation problem of a polymer under an external applied stretching force has received renewed attention [14-27]. In ref. [27], the authors studied the 2D collapse transition of a polymer under external stretching force, through a partially directed lattice model. In this model, both the flexible ($\Delta=0$) and the semi-flexible ($\Delta > 0$) cases can be studied analytically. In this paper we extend the method of ref [27] to study the partially directed polymer near an attractive wall in two dimensions. Such situation can in principle occur in biophysics for a semi-flexible polymer such as DNA to be close to an attractive membrane. We are able to obtain analytic result for this model. We find that for wall attraction $\varepsilon_1$ smaller than the non-sequential nearest neighbor attraction $\varepsilon$, the fraction of monomers at the wall is zero and the model is the same as that of a polymer without a wall. However, for $\varepsilon_1$ greater than $\varepsilon$, the fraction of monomers at the wall undergoes a first order transition from unity at low temperature and force, to zero at higher temperatures and forces. We present phase diagram for this transition. Our results are confirmed by Monte-Carlo simulations.

In section II we introduce the model. In section III we describe the method used in solving it and the results obtained. Section IV is the Summary.

## II. The Model

The 2D partially directed polymer of N identical units on a square lattice is shown in Fig. 1. The left end of the polymer is attached to an impenetrable wall, shown in the figure as the y-axis. The model is partially directed in the sense that monomer units can be added only in the positive x-direction, while in the transverse direction, they can be added both in the positive and negative y-directions. The length of the bond connecting two consecutive monomers i and i+1 is fixed to $a_0$. If any two monomer i and i+m($m \geq 3$) occupy nearest neighboring lattice sites, an attractive energy of magnitude $\varepsilon$ is gained. Usually real polymers are semi-flexible. We associate an energy penalty of magnitude $\Delta$ to each local direction change of the chain [22, 28]. If a monomer happens to be on the wall, an attractive energy of magnitude $\varepsilon_1$ is gained. In addition, the other end of the polymer is pulled by a stretching force f in the positive x direction. With this stretching force, this partially directed model is not that unphysical since it makes the monomers tend to align in the positive x direction.

## III. The Method and Results

The partially directed model described above can be solved analytically using the method of [27], which we will now follow closely. To calculate the free energy density of the polymer, a given configuration of the 2D chain is divided into a linear sequence of β-sheet segments and coil segments using the Lifson approach [27, 29]. A β-sheet segment is defined as a folded segment of $n_\beta \geq 2$ consecutive columns, in which contacting interactions exist between any two adjacent ones. Two consecutive β-sheet segments are separated by a coil segment, which is a segment of $n_c \geq 0$ consecutive columns in which all monomers are free of contacts. For example, the configuration shown in Fig. 1 consists of one β-sheet of 22 monomers and one coil of five monomers. After making such a distinction between β-sheets and coils, we proceed by first calculating the partition functions of β-sheets and coils separately.

Under the action of an external force f, the energy of a β-sheet of $n_\beta$ columns is

$$E_{b\beta} = -\varepsilon \sum_{j=1}^{n_\beta - 1} v(l_j, l_{j+1}) + 2(n_\beta - 1)\Delta - \varepsilon_1 l_1 - n_\beta f a_0 \qquad (1)$$

where $\varepsilon$ is attractive energy between the nearest neighboring lattice sites, $\varepsilon_1$ is the energy gain of each monomer attached on the wall. $\Delta$ is the energy penalty of each local direction change of the chain, $l_j$ is the number of monomers in the j-th column of the β sheet, and $v(l_j, l_{j+1}) = \min(l_j, l_{j+1}) - 1$. The partition function of a β sheet with $n \geq 4$ monomers is

$$Z_{b\beta}(n) = \sum_{n_\beta=2}^{[n/2]} \sum_{\{l_j \geq 2\}} \delta^n_{l_1+\cdots+l_{n_\beta}} b^{l_1} p^{n_\beta} s^{2(n_\beta-1)} \prod_{j=1}^{n_\beta-1} a^{v(l_j, l_{j+1})} \qquad (2)$$

where $a = e^{\varepsilon/T}$, $b = e^{\varepsilon_1/T}$, $p = e^{fa_0/T}$, $s = e^{-\Delta/T}$, and T is the temperature. It is easier to calculate the generating function $G_{b\beta}(\zeta)$ of the partition function $Z_{b\beta}(n)$ than to calculate $Z_{b\beta}(n)$ directly.

$$G_{b\beta}(\zeta) \equiv \sum_{n=4}^{\infty} (\zeta/a)^n Z_{b\beta}(n)$$

$$= \sum_{n=4}^{\infty} \sum_{m=2}^{[n/2]} \sum_{\{l_j \geq 2\}} \delta^n_{l_1+l_2+\cdots l_m} s^{2(m-1)} p^m b^{l_1} \left(\frac{\zeta}{a}\right)^{\frac{l_1}{2}} \left(\frac{\zeta}{a}\right)^{\frac{l_m}{2}} \prod_{i=1}^{m-1}\left[\left(\frac{\zeta}{a}\right)^{\frac{l_i}{2}} a^{\min(l_i, l_{i+1})-1} \left(\frac{\zeta}{a}\right)^{\frac{l_{i+1}}{2}}\right]$$

(3)

The configurational energy of a coil segment with a wall is

$$E_{bc} = m_c \Delta - \varepsilon_1 l_1 - n_c f a_0 \qquad (4)$$

where $m_c$ is the total number of bends in the configuration, $l_1$ is the number of monomers on the wall. To calculate the partition function $Z_{bc}(n)$ of a coil segment of n monomers, one needs to distinguish among four different boundary conditions.

(i) Both the wall and the right-most column contain only one monomer. The partition function for such a situation is denoted as $Z_{bc}^{1,1}(n,n_c)$.

(ii) The wall contains only one monomer, while the right-most column contains two or more monomers. The partition function for such a situation is denoted as $Z_{bc}^{1,2}(n,n_c)$

(iii) The wall contains two or more monomers, while the right-most column contains just one monomer. The partition function for such a situation is denoted as $Z_{bc}^{2,1}(n,n_c)$.

(iv) Both the wall and the right-most column contain two or more monomers. The partition function for such a situation is denoted as $Z_{bc}^{2,2}(n,n_c)$.

We can write down the following iteration equations for the four partition functions:

$$Z_{bc}^{1,1}(n,n_c) = pZ_{bc}^{1,1}(n-1,n_c-1) + psZ_{bc}^{1,2}(n-1,n_c-1), (n \geq 2, 2 \leq n_c \leq n) \quad (5)$$

$$Z_{bc}^{1,2}(n,n_c) = Z_{bc}^{1,2}(n-1,n_c) + 2psZ_{bc}^{1,1}(n-2,n_c-1) + ps^2 Z_{bc}^{1,2}(n-2,n_c-1), (n \geq 3, 2 \leq n_c \leq n-1) \quad (6)$$

$$Z_{bc}^{2,1}(n,n_c) = pZ_{bc}^{2,1}(n-1,n_c-1) + psZ_{bc}^{2,2}(n-1,n_c-1), (n \geq 2, 2 \leq n_c \leq n-1) \quad (7)$$

$$Z_{bc}^{2,2}(n,n_c) = Z_{bc}^{2,2}(n-1,n_c) + 2psZ_{bc}^{2,1}(n-2,n_c-1) + ps^2 Z_{bc}^{2,2}(n-2,n_c-1). (n \geq 4, 2 \leq n_c \leq n-2) \quad (8)$$

The initial conditions are: $Z_{bc}^{1,1}(1,n_c) = pb\delta_{n_c}^1$; $Z_{bc}^{1,2}(1,n_c) = Z_{bc}^{1,2}(2,n_c) = 0$;

$$Z_{bc}^{2,1}(1,n_c) = Z_{bc}^{2,1}(2,n_c) = 0; \text{ and } Z_{bc}^{2,2}(1,n_c) = 0, Z_{bc}^{2,2}(2,n_c) = pb^2\delta_{n_c}^1. \quad (9)$$

Except for the initial conditions in (9), these equations are the same as those of ref. [27] for the case without a wall.

By iterating the above equations, the generating functions $G^{i,j}{}_{bc}(\zeta)$ of $Z^{i,j}{}_{bc}(n,n_c)$, i,j =1 or 2 defined as

$$G^{i,j}{}_{bc}(\zeta) = \sum_{n=1}^{\infty} \left(\frac{\zeta}{a}\right)^n \sum_{n_c=1}^{n} Z^{i,j}{}_{bc}(n,n_c) \quad (10)$$

can be calculated as:

$$G_{bc}^{1,1}(\zeta) = \frac{\zeta pb(a^2 - \zeta a - \zeta^2 ps^2)}{B(\zeta,a,p)} \quad (11)$$

$$G_{bc}^{1,2}(\zeta) = \frac{2b\zeta^3 p^2 s}{B(\zeta, a, p)} \qquad (12)$$

$$G_{bc}^{2,1}(\zeta) = \frac{\zeta^3 p^2 s b^2 (a-\zeta)}{(a-b\zeta)B(\zeta, a, p)} \qquad (13)$$

$$G_{bc}^{2,2}(\zeta) = \frac{pb^2\zeta^2(a-\zeta)(a-p\zeta)}{(a-b\zeta)B(\zeta, a, p)} \qquad (14)$$

where
$$B(\zeta, a, p) = a^3 - \zeta a^2(1+p) + \zeta^2 a p(1-s^2) - \zeta^3 p^2 s^2 \qquad (15)$$

One can easily check that the above equations reduce to those of ref. [27] in the case b=1, when there is no interaction with the wall. By Taylor expanding the $G^{i,j}{}_{bc}(\zeta)$ in $\zeta$ one can check that the results agree with those obtained by solving equations (5)-(10) to all orders in $\zeta$, so that they are indeed exact solutions.

The total partition function for a coil segment of n monomers attached to a wall is

$$Z_{bc}(n) = s\sum_{n_c=1}^{n} Z_{bc}^{1,1}(n, n_c) + s^2 \sum_{n_c=2}^{n-1} Z_{bc}^{1,2}(n, n_c) + s\sum_{n_c=2}^{n-1} Z_{bc}^{2,1}(n, n_c) + s^2 \sum_{n_c=2}^{n-1} Z_{bc}^{2,2}(n, n_c) \qquad (16)$$

while $Z_{bc}(0) \equiv 0$. The factors s and $s^2$ in (16) come from the extra bends or corners that arise when the coil partition function with a wall is connected to a $\beta$-sheet partition function without a wall. The generating function for the coil segment partition function is

$$G_{bc}(\zeta) = \sum_{n=0}^{\infty} \left(\frac{\zeta}{a}\right)^n Z_{bc}(n)$$
$$= sG_{bc}^{1,1}(\zeta) + s^2 G_{bc}^{1,2}(\zeta) + sG_{bc}^{2,1}(\zeta) + s^2 G_{bc}^{2,2}(\zeta) \qquad (17)$$

The generating function for the coil segment partition function of a polymer without wall was obtained in ref. [27]. It can be obtained from our equations above by simply setting b=1.

In the lattice model, any configuration of the polymer is a chain of some monomers either $\beta$-sheet $G_{b\beta}$, or random coil $G_{bc}$ which are attached to the wall. These parts can then be attached to polymer parts which are not attached to a wall, either $G_\beta$ or $G_c$, which can be obtained either from ref [27] or from equations (10)-(17) by setting b=1. The part of polymer which is not attached to the wall has the two types of segments occurring alternately along the polymer. Since every configuration of the polymer is of the form

$b-\beta-c-\beta-c\cdots$ or $b-c-\beta-c-\beta\cdots$ the generating function of the total partition function is

$$G_b(\zeta) = \sum_{N=0}^{\infty} \left(\frac{\zeta}{a}\right)^N Z_b(N)$$
$$= G_{b\beta} + G_{bc} + G_{bc}G_\beta G_c + G_{bc}(G_\beta G_c)^2 + G_{bc}(G_\beta G_c)^3 + \cdots$$
$$+ G_{b\beta}G_c + G_{b\beta}G_c G_\beta G_c + G_{b\beta}G_c(G_\beta G_c)^2 + \cdots$$
$$G_{b\beta}G_c G_\beta + G_{b\beta}(G_c G_\beta)^2 + \cdots$$
$$+ G_{bc}G_\beta + G_{bc}G_\beta G_c G_\beta + G_{bc}G_\beta(G_c G_\beta)^2 + \cdots$$

$$= G_{b\beta} + G_{bc} + \frac{G_{bc}G_\beta(1+G_c) + G_{b\beta}G_c(1+G_\beta)}{1-G_\beta(\zeta)G_c(\zeta)} \quad (18)$$

Note that this result reduces to that of ref [27] for the case b=1, when $G_{bc}$ and $G_{b\beta}$ reduce to $G_c$ and $G_\beta$ respectively.

The generating function $G_\beta$ is given by [27]

$$G_\beta(\zeta) = \frac{p^2 s^2}{a} \sum_{i,j,k=1} \frac{x_i x_j A_{ik}(\zeta) A_{jk}(\zeta) \lambda_k(\zeta)}{1-(ps^2/a)\lambda_k(\zeta)} \quad (19)$$

In Eqn. (19), $x_j = (\zeta/a)^{(j+1)/2}$; $\lambda_1(\zeta) \geq \lambda_2(\zeta) \geq \cdots$ are the eigenvalues of a LxL real-symmetric matrix $\Lambda(\zeta)$ with elements $\Lambda_{ij}(\zeta) = \zeta^{1+(i+j)/2}/a^{|i-j|/2}$ $(i,j=1,2,\cdots L)$; and the orthogonal matrix $A(\zeta)$ contains the eigenvectors of matrix $\Lambda(\zeta)$. The parameter L should be infinity. When $\zeta > 1, \lambda_1(\zeta) = +\infty$, and consequently $G_\beta(\zeta)$ is not properly defined. When $\zeta \leq 1$, all the eigenvalues of matrix $\Lambda(\zeta)$ are finite, and the value $G_\beta(\zeta)$ can be calculated by Eqn. (19).

There are three singular points of Eqn.(18) given by the divergences of $G_{b\beta}(\zeta), G_{bc}(\zeta)$ and the vanishing of the denominator $1-G_\beta(\zeta)G_c(\zeta)$. We will first determine the value $\zeta_1$ at which $G_{b\beta}(\zeta_1)$ diverges. From Eqn. (2)

$$G_{b\beta}(\zeta) = \sum_{m=2}^{\infty} s^{2m-2} p^m \sum_{\{l_j \geq 2\}} b^{l_1} \left(\frac{\zeta}{a}\right)^{\frac{l_1}{2}} \left(\frac{\zeta}{a}\right)^{\frac{l_m}{2}} \left(\frac{\zeta}{a}\right)^{\frac{l_1}{2}} a^{\min(l_1,l_2)-1} \left(\frac{\zeta}{a}\right)^{\frac{l_2}{2}} \prod_{i=2}^{m-1} \left(\frac{\zeta}{a}\right)^{\frac{l_i}{2}} a^{\min(l_i,l_{i+1})-1} \left(\frac{\zeta}{a}\right)^{\frac{l_{i+1}}{2}}$$

$$= \sum_{m=2}^{\infty} s^{2m-2} p^m \sum_{\{l_j \geq 2\}} \left(b\frac{\zeta}{a}\right)^{l_1} a^{\min(l_1,l_2)-1} \left(\frac{\zeta}{a}\right)^{\frac{l_2}{2}} \left(\frac{\zeta}{a}\right)^{\frac{l_m}{2}} \prod_{i=2}^{m-1} \left(\frac{\zeta}{a}\right)^{\frac{l_i}{2}} a^{\min(l_i,l_{i+1})-1} \left(\frac{\zeta}{a}\right)^{\frac{l_{i+1}}{2}}$$

The sum over $l_1$ can be written as

$$\sum_{l_1=2}^{\infty}\left(b\frac{\zeta}{a}\right)^{l_1} a^{\min(l_1,l_2)} = \sum_{l_1=2}^{l_2}\left(b\frac{\zeta}{a}\right)^{l_1} a^{l_1} + \sum_{l_1=l_2+1}^{\infty}\left(b\frac{\zeta}{a}\right)^{l_1} a^{l_2}$$

Clearly the second sum diverges at $\frac{b\zeta_1}{a} = 1$ or $\zeta_1 = a/b$. From Eqns.(12) to (15) one can see that this gives also the divergence of $G_{bc}(\zeta)$. The singular point $\zeta_0$ coming from the vanishing of the denominator $1 - G_\beta(\zeta)G_c(\zeta)$ is independent of b.

The fraction of monomers at the wall is given by

$$<n> = -\frac{\partial \ln \zeta_i}{\partial \ln b}, i = 0,1 \qquad (20)$$

Similarly, the relative extension of the polymer is given by

$$<l> = -\frac{\partial \ln \zeta_i}{\partial \ln p}, i = 0,1., \qquad (21)$$

Therefore the fraction of monomer is either one or zero depending on whether $\zeta_1$ or $\zeta_0$, respectively, is used in Eqn. (20). Since the physical values of $\zeta$ are given by $\zeta \leq 1$, the condition $\zeta_1 = a/b = \exp((\varepsilon - \varepsilon_1)/T)$ can only be satisfied by $\varepsilon_1 \geq \varepsilon$. For $\varepsilon_1 < \varepsilon$, the only singular point of Eqn.(18) is $\zeta_0$, independent of b. This will give the fraction of monomer equal to zero according to Eqn. (20) and the wall has no effect on the polymer. The relative extension of the polymer is then given by Eqn. (21) using $\zeta_0$. This will give the same result obtained in [27]. In the following we will only consider the case $\varepsilon_1 \geq \varepsilon$. In this case the free energy per monomer $\zeta_1$ is small at low temperatures whereas the free energy per monomer $\zeta_0$ is known to be large at low temperatures [27]. Therefore, at low temperatures and for $\varepsilon_1 \geq \varepsilon$, the free energy per monomer in the $\beta$-state $\zeta_1$ is lower than $\zeta_0$ and the polymer is in the $\beta$-state. Using $\zeta_1$ in Eqns.(20) and (21), one find that the fraction of monomers at the wall is unity and the relative extension is zero, since $\zeta_1$ is independent of p. The polymer is completely adsorbed at the wall. As the temperature increases, $\zeta_1$ increases and approaches unity, while $\zeta_0$ decreases and eventually becomes negative [27]. Therefore there always exists a temperature above which $\zeta_0$ is smaller than $\zeta_1$ and $\zeta_0$ then takes over as the physical chemical potential per monomer. At this point, according to Eqn. (20), the fraction of monomers at the wall is zero and the polymer is desorbed from the surface. The relative extension is then obtained by using $\zeta_0$ in Eqn. (21), yielding the same result as in [27]. At fixed temperature T, the phase diagram can therefore be obtained by substituting $\zeta = a/b = \exp((\varepsilon - \varepsilon_1)/T)$ into the equation $1 - G_\beta(\zeta,T,f)G_c(\zeta,T,f) = 0$ and solving for the critical force f. Such phase diagrams for Δ=0 and Δ=0.25 are presented in Figures 2a and 2b respectively. In both figures, for each value of $\varepsilon_1$, the region above (below) the curve represents desorbed (adsorbed) phase, respectively. In Figure 3 we present the phase diagram for the particular case of no external force. In this figure, for each value of $\Delta$, the region above (below) the curve represents absorbed (desorbed) phase respectively. Since our wall is one dimensional, the desorption transition of a self-avoiding polymer from this wall is very similar to the DNA denaturation transition (if driven by temperature) or the DNA unzipping transition (if driven by force). Both transitions are found to be first-order by various earlier theoretical and experimental studies. We have studied a different problem here, i.e., the competition between wall adsorption and formation of beta-sheet structures. But as we showed in our work, when the adsorption energy is large enough, the transition will also be first order.

In order to support our theoretical result we have performed Monte-Carlo simulation of our model, using the Monte-Carlo method of ref. [27]. The simulations are performed for a polymer of length N=4900, at fixed wall attraction $\varepsilon_1 = 1.2\varepsilon$. For both cases of flexible ($\Delta = 0$) and semiflexible ($\Delta = 0.25\varepsilon$) we have calculated the extension versus temperature curve at zero external force and the extension versus force curve at fixed temperature T=ε. In Figure 4a we have plotted the extension versus temperature at zero external force, at wall attraction $\varepsilon_1 = 1.2\varepsilon$. We can see that the extension is zero for temperature below about 1.1 ε for stiffness $\Delta = 0$ and jumps discontinuously to a finite value at higher temperatures. For stiffness $\Delta = 0.25\varepsilon$, the extension is zero for temperature below T=1.3 ε and jumps discontinuously to a finite value at higher temperatures. This is consistent with our theoretical result in Figure 3. In Figure 4b we have plotted the extension versus force at fixed temperature T=ε, at wall attraction $\varepsilon_1 = 1.2\varepsilon$. Again the simulation results are consistent with the theoretical results in Figure 2a and 2b. The error bars in Figures 4a and 4b are smaller than the size of the data points.

## IV. Summary

In summary we have studied analytically a model of a two dimensional, partially directed, flexible or semiflexible polymer, attached to an attractive wall which is perpendicular to the preferred direction. In addition, the polymer is stretched by an externally applied force. We find that for wall attraction $\varepsilon_1$ smaller than the non-sequential nearest neighbor attraction $\varepsilon$, the fraction of monomers at the wall is zero and the model is the same as that of a polymer without a wall. However, for $\varepsilon_1$ greater than $\varepsilon$, the fraction of monomers at the wall undergoes a first order transition from unity at low temperature and force, to zero at higher temperatures and forces. We present phase diagram for this transition. Our results are confirmed by Monte-Carlo simulations.

Acknowledgement. Research supported by the Louisiana Board of Regents Support Fund Contract Number LEQSF(2007-10)-RD-A-29 and by China 973-Project Grant 2007CB935903. PML thanks the Institute of Theoretical Physics, Chinese Academy of Sciences for hospitality.

# References

[1] G.J. Fleer, M.A. Cohen-Stuart, J.M.H.M. Scheutjens, T. Grosgrove, B. Vincent, *Polymers at Interface,* Chapman Hall, London, 1993
[2] P.G. de Gennes, J. Phys. (Paris), **37**, 1445 (1976)
[3] P.G. de Gennes, Macromolecules **14**, 1637 (1981)
[4] P.G. de Gennes, Macromolecules, **13**, 1069 (1980)
[5] P.G. de Gennes, P. Pincus, J. Phys. Lett.(Paris), **44**, L-241 (1983)
[6] P.G. de Gennes, Adv. Colloid Inerface Sci. **27**, 189 (1987)
[7] E. Eisenriegler, K. Kremer, K. Binder, J. Chem. Phys. **77**, 6296 (1982)
[8] R. Descas, J.-U. Somer, A. Blumen, J. Chem. Phys. **120**,8831 (2004)
[9] S. Metzger, M. Müller, K. Binder, J. Baschnagel, Macromol. Theory Simul. **11**, 985 (2002)
[10] A. Milchev, K. Binder, Macromolecules, **29**, 343 (1996)
[11] A. Takahashi, M. Kawaguchi, Adv. Polym. Sci. **46**, 1 (1982)
[12] M. Cohen-Stuart, T. Cosgrove, B. Vincent, Adv. Colloid Interface Sci. **24**, 143 (1985)
[13] For a brief review, see Bustamante, C.; Bryant, Z. and Smith, S.B., *Nature* **421**, 423 (2003)
[14] T. Strick, J.F. Allemand, V. Croquette and D. Bensimon, Physics Today **54**, 46 (2001)
[15] C. Danilowicz, C.H. Lee, V.W. Coljee and M. Prentiss, Phys. Rev. **E75**, 030902(R) (2007)
[16] H. Clausen-Schaumann, M. Seitz, R. Krautbauer and H.E. Gaub, Cruu. Opin. Chem. Biol. **4**, 524 (2000)
[17] C. G. Baumann, V.A. Bloomfield, S. B. Smith, C. Bustamante, M.D. Wang and S.M. Block, Biophys. J. **78**, 1965 (2000)
[18] A. Halperin and E.B. Zhulina, Europhys. Lett. **15**, 417 (1991)
[19] P. Grassberger and H.P. Hsu, Phys. Rev. **E65**, 031807 (2002)
[20] D. Marenduzzo, A. Maritan, A. Rosa and F. Seno, Phys. Rev. Lett. **90**, 088301 (2003)
[21] A. Rosa, D. Marenduzzo, A. Maritan and F. Seno, Phys. Rev. **E67**, 041802 (2003)
[22] S. Kumar and D. Giri, Phys. Rev. **E72**, 052901 (2005)
[23] E.Orlandini, M. Tesi and S. Whittington, J. Phys. **A37**, 1535 (20045)
[24] S. Kumar, I. Jensen, J.L. Jacobson and A. J. Guttmann, Phys. Rev. Lett. **98**, 128101 (2007)
[25] S. Kumar and G. Mishra, Phys. Rev. **E78**, 011907 (2008)
[26] J. Krawczyk, I. Jensen, A. L. Owczrak, and S. Kumar, Phys.Rev. E **79**, 031912 (2009).

[27] H. Zhou, Jie Zhou, Z.-C. Ou-Yang and S. Kumar, Phys. Rev. Lett. **97**, 158302 (2006)
[28] S. Doniach, T. Garel and H. Orland, J. Chem. Phys. **105**, 1601 (1996)
[29] S. Lifson, J. Chem. Phys. **40**, 3705 (1964)

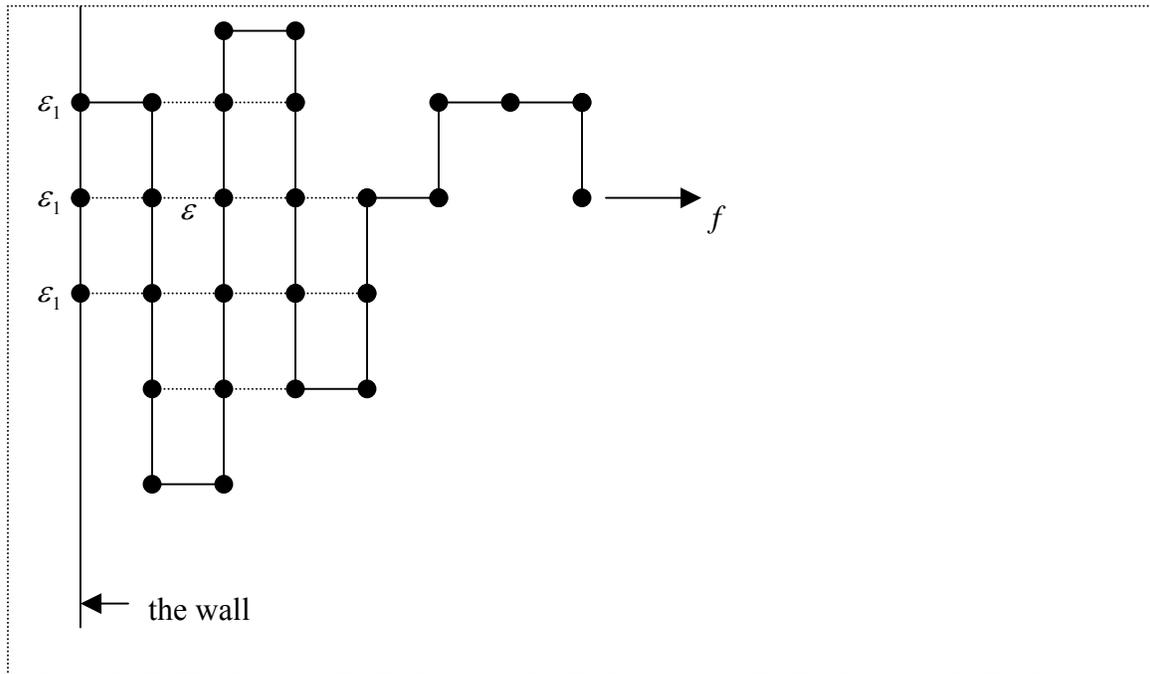

Figure 1: Configuration of a polymer attached to an attractive wall. Solid dots represent monomers. Dashed lines represent attraction $\varepsilon$. The y-axis on the left denotes the wall.

Monomers at the wall experience an extra attraction $\varepsilon_1$. The right end of the polymer is subjected to an external force f.

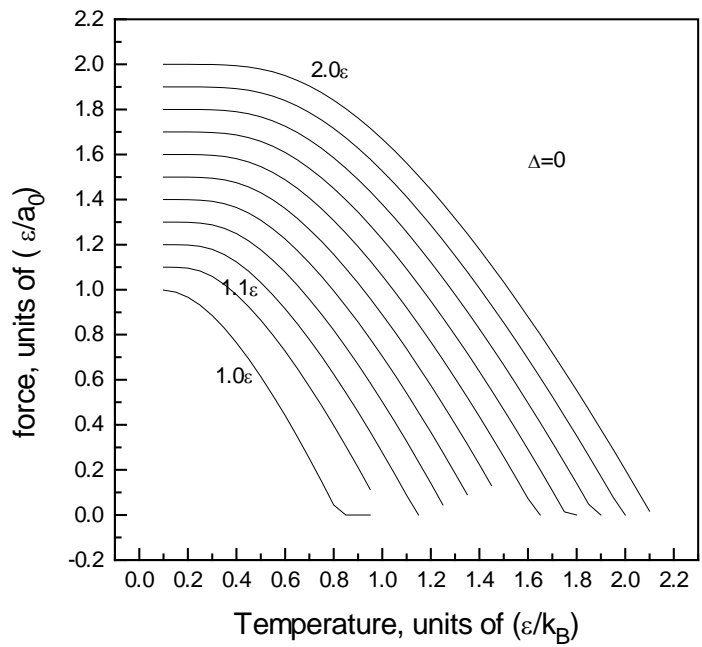

Figure 2a: Phase diagram for flexible polymer (Δ=0), at different wall attraction $\varepsilon_1 = 1.0\varepsilon, 1.1\varepsilon, \cdots, 2\varepsilon$.

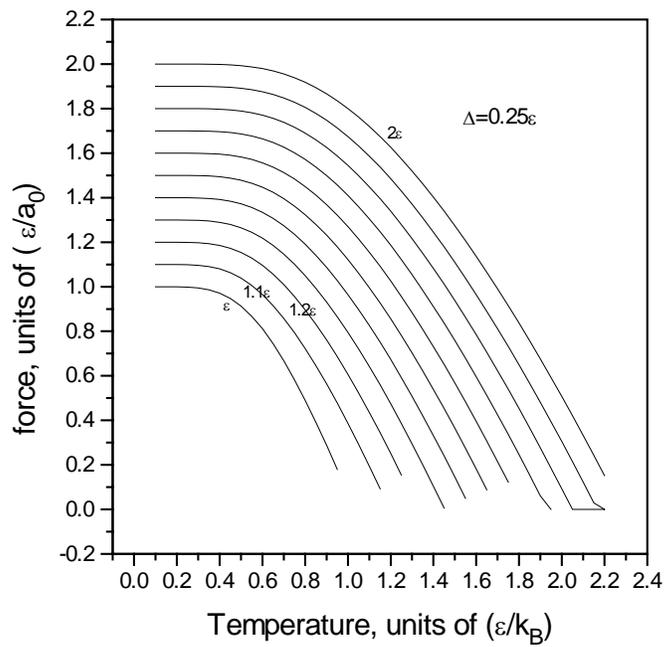

Figure 2b: Phase diagram for semiflexible polymer ($\Delta=0.25\varepsilon$) at wall attraction $\varepsilon_1 = 1.0\varepsilon, 1.1\varepsilon, \cdots, 2\varepsilon$.

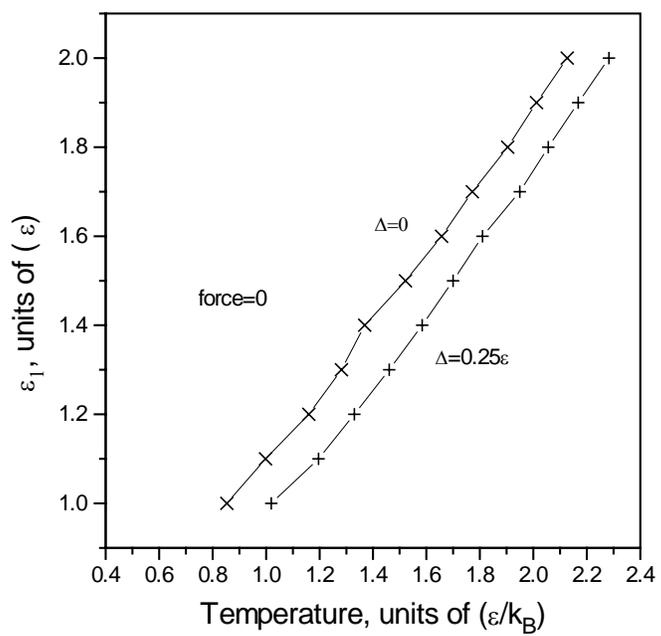

Figure 3: Phase diagram at zero external force, for flexible (Δ=0) and semiflexible (Δ=0.25ε).

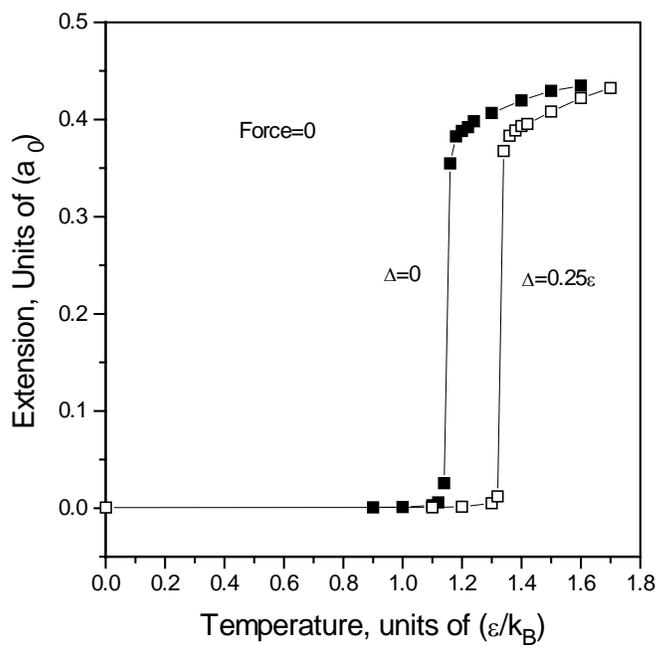

Figure 4a: Extension versus temperature at zero external force for flexible ($\Delta = 0$) and semiflexible ($\Delta = 0.25\varepsilon$) polymers, at wall attraction $\varepsilon_1 = 1.2\varepsilon$.

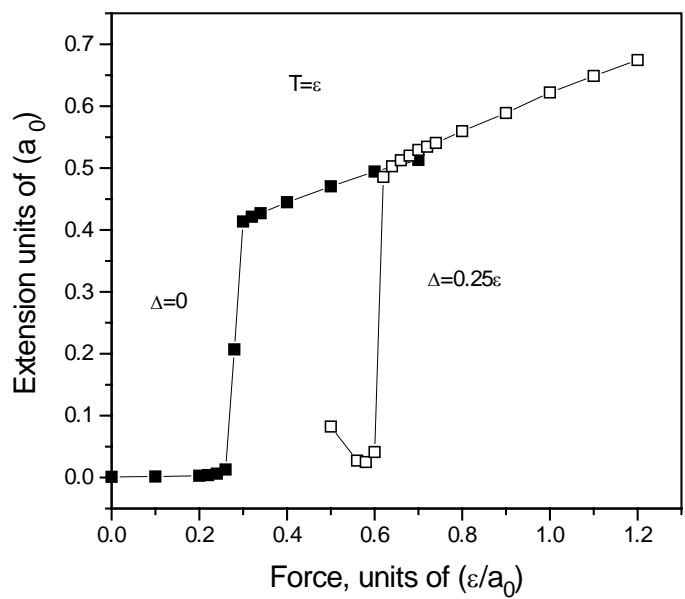

Figure 4b: Extension versus force at fixed temperature T=$\varepsilon$ for flexible($\Delta = 0$) and semiflexible ($\Delta = 0.25\varepsilon$) polymers, at wall attraction $\varepsilon_1 = 1.2\varepsilon$.